# Simultaneous electronic and the magnetic excitation of a ferromagnet by intense THz pulses

Mostafa Shalaby[1], Carlo Vicario[1], and Christoph P. Hauri[1,2]

[1]Paul Scherrer Institute, SwissFEL, 5232 Villigen PSI, Switzerland.

[2]École Polytechnique Fédérale de Lausanne, 1015 Lausanne, Switzerland.

*Correspondence to: most.shalaby@gmail.com, carlo.vicario@psi.ch, and christoph.hauri@psi.ch

*The speed of magnetization reversal is a key feature in magnetic data storage. Magnetic fields from intense THz pulses have been recently shown to induce small magnetization dynamics in Cobalt thin film on the sub-picosecond time scale. Here, we show that at higher field intensities, the THz electric field starts playing a role, strongly changing the dielectric properties of the cobalt thin film. Both the electronic and magnetic responses are found to occur simultaneously, with the electric field response persistent on a time scale orders of magnitude longer than the THz stimulus.*

The speed of magnetization reversal is a key feature for ultrafast magnetic storage technology [1-2]. In the state of the art data recording technology, the magnetization reversal, i.e., the complete inversion of the direction of the magnetization vector (**M**) occurs on a relatively long time scale on the order of the nanosecond. In the past, femtosecond optical pulses have been used to induce faster magnetic phase transition, but the impact of the optical pulse leads to thermal effects and therefore to a long recovery time (nanoseconds), which significantly limits the magnetic writing rate [3-6]. Using intense magnetic field pulses from a relativistic electron beam, magnetic switching on a picosecond timescale was observed post mortem [1-2]. However, time-resolved exploration of the interplay of the electric and the magnetic field component with the ferromagnetic sample could not be performed at this large scale facility.

An alternative approach towards precessional magnetic switching on the picosecond timescale relies on the magnetic field component of a strongly asymmetric single cycle terahertz (THz) pulse phase-locked to the spins [7-15]. As the THz pulse carrier frequency matches well the natural timescale of the spin motions direct control of the spin dynamics with the THz magnetic field component becomes feasible by inducing a torque [16-17]. Recently, this concept has been proven experimentally by *Vicario et al* where a small excursion of the spins was introduced non-resonantly by an intense 0.4 Tesla THz field [17]. The observed coherent spin dynamics were more than an order of magnitude faster than in previous studies. Effects associated to the THz electric field were not observed. While this proof of principle experiment showed the great potential of intense THz radiation for precessional spin motion control, full magnetic switching requires much higher field intensities than what was available in ref. [17]. Here, we report on dynamics induced by an up-scaled THz pulse intensity targeting large amplitude magnetization dynamics. We observed that at high field intensities, the THz electric field starts playing a significant role in changing the dielectric properties of the magnetic material.

The THz source used for the present studies provides intense THz pulses with maximum electric and magnetic field strengths of 6.7 MV/cm and 2.23 T, respectively. The corresponding peak intensity is 59 GW/cm$^2$. The THz was generated using optical rectification of a short wavelength infrared pulses ($\lambda_c$=1.5 μm) in an organic crystal DSTMS [18-19]. Figure 1a and 1b show the temporal field profile recorded by electro-optical sampling and the corresponding spectral amplitude. The THz spot in the focus, recorded using an uncooled micro-bolometer array (NEC Inc.) is shown in Fig. 1c.

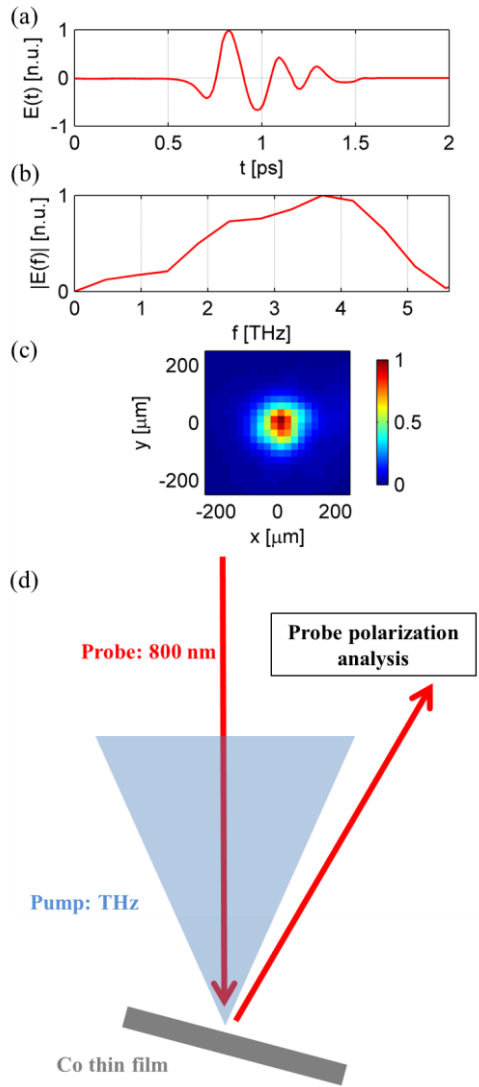

FIG. 1. (a) The temporal trace of the exciting THz pulse retrieved using electro-optical sampling technique (peak of 6.7 MV/cm). (b) The corresponding amplitude spectrum. (c) The THz spot size at the sample position measured using a micro-bolometer array-based camera. (d) Schematic diagram of the beams configurations on the sample.

To measure the THz-induced nonlinear dynamics, we used a collinear THz-pump/optical Kerr effect (KE) probe scheme. The investigated sample is a 20 nm-thick Co film deposited on a Si substrate and capped with 2 nm-thick layer of Pt. The latter is a protection layer which plays no role in the experiment presented here. We verified that by comparing the measurement against uncoated fresh sample. The beam configuration at the sample position is shown in Fig. 1d: the THz pump and a collinear 800-nm centered probe beam impinge on the sample at an angle of 15° measured from the normal. The reflected probe beam was then collected and the THz-induced KE rotation was analyzed using a balanced detection scheme (a quarter wave plate followed by a Wollaston prism). An external magnetic field **B** bias (parallel to the sample plane and the plane of incidence) is used to prepare a well-defined in-plane magnetization state **M** prior to pumping. During the measurements the THz polarization direction could be altered by rotating the the THz generation crystal along with the near infrared pump beam polarization.

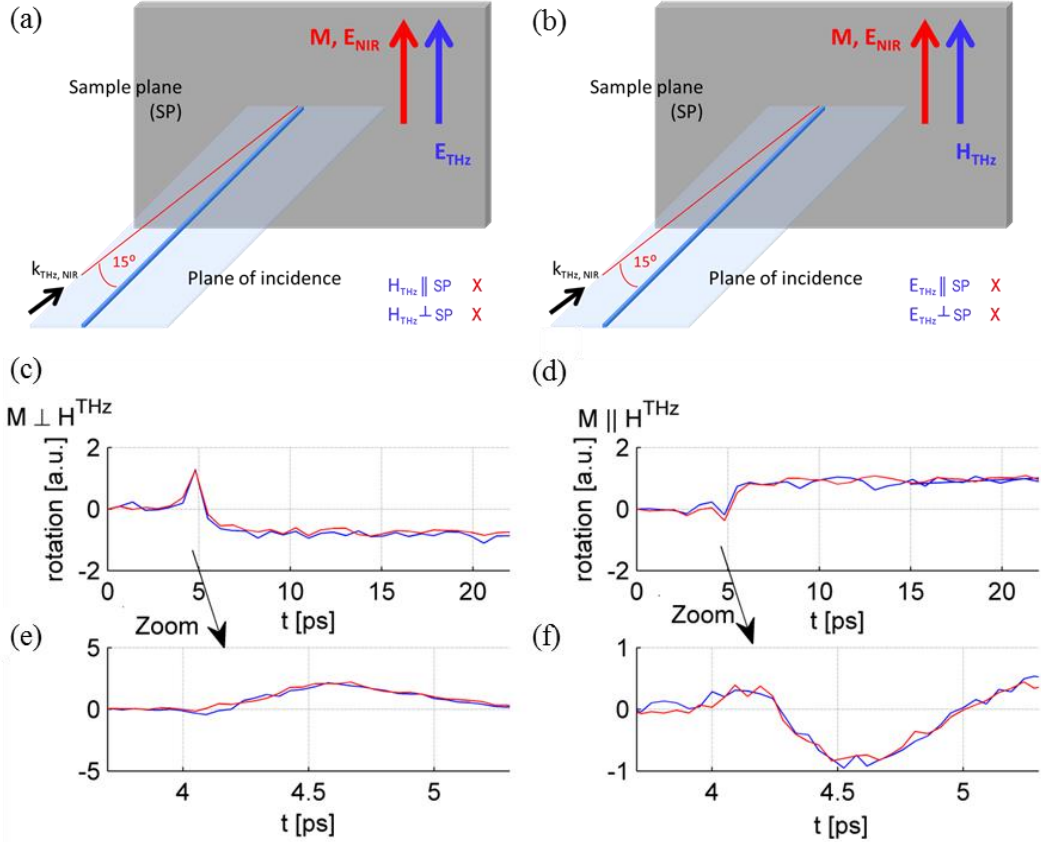

*FIG. 2 The experimental layout in (a) the precession $\pm M \perp H^{THz}$ and (b) non-precession $\pm M \parallel H^{THz}$ configurations. (c) and (d) show the corresponding THz-induced electronic and magnetic dynamics (optical probe rotation). The transient curves are zoomed in and plotted in (e) and (f). The red and blue curves refer to the positive and negative directions of the applied static field (**B**) and thus magnetization (**M**).*

Nonthermal temporal magnetic evolution under the application of an external (THz) magnetic field is macroscopically governed by the Landau-Lifschitz-Gilbert (LLG) equation:

$$\frac{\partial \mathbf{M}}{\partial t} = -|\gamma| \mathbf{M} \times \mathbf{H}_{\text{eff}} + \frac{\alpha}{M_s}\left(\mathbf{M} \times \frac{\partial \mathbf{M}}{\partial t}\right) \qquad (1)$$

where $\gamma$, $\alpha$ and $M_s$ are the gyromagnetic ratio, Gilbert damping coefficient, and material-dependent saturation magnetization, respectively. $H_{\text{eff}}$ is the effective field summing the contributions from the internal and external (here THz, $\mathbf{H}^{THz}$) fields. An intense enough $\mathbf{H}^{THz}$ (with a nonzero angle to the direction of **M**) perturbs this alignment driving time-dependent dynamics in **M**. The realignment process consists of two dynamics, described by the two corresponding terms in the LLG. This includes ultrafast precessional dynamics taking place on the THz (stimulus) time scale and the damping mechanism which tries to orient **M** towards the new direction of equilibrium $\mathbf{H}_{eff}$ on a much longer time scale (governed by the material parameters). The damping intensity is generally much weaker than the ultrafast precessions. At relatively weak excitation, the amplitude of the induced ultrafast precessions is nearly a linear function of $\mathbf{H}^{THz}$.

The simplified LLG model (1) does not take into account the effect of the THz electric field and thus may be inappropriate for the physics observed at high fields to describe. High field intensities (on the order of several Tesla in the sub-THz range and much more as the excitation frequency increases [20-21]), however, are required for Terahertz-induced magnetization reversal. In this article we were aiming to increase the excitation intensity in order to study the role of the THz electric field component ($\mathbf{E}^{THz}$) in the observed dynamics. Two major $\mathbf{E}^{THz}$-induced effects can be considered. The first mechanism is the change in the dielectric tensor due to Kerr-like and thermal nonlinearities. The second effect is the thermal demagnetization of the sample induced by heating the electrons above the Curie temperature, similar to the majority of laser-induced magnetization dynamics in the optical regimes. The second mechanism depends on the excitation fluence. In our experimental setup, the sample is exposed to two external excitations, namely the THz (electric and magnetic) fields and the bias magnetic field. In phase-sensitive detection as we used here, one of the external excitations is modulated at a frequency which is used as a reference for our acquisition system. This allows us to distinguish between the magnetic and non-magnetic (electronic) dynamics. While modulating the THz pump reveals both dynamics, modulating the external bias shows on the magnetic ones.

The nonlinear dynamics induced by $\mathbf{E}^{THz}$ and $\mathbf{H}^{THz}$ are very different. For simplicity, we assume no coupling between $\mathbf{E}^{THz}$ and **M**. This assumption is manifested by the fact that changes in **M** require a nonzero torque $\mathbf{M} \times \mathbf{H}_{eff}$ (i.e. it vanishes for $\mathbf{M} \parallel |\mathbf{H}^{THz}|$ and is a maximum for $\mathbf{M} \perp |\mathbf{H}^{THz}|$) and that the excitation of the electronic system does not depend on the direction of **M**. Second, while the $\mathbf{H}^{THz}$-induced changes in the experimentally-dominant out-of-plane **M** component are generally linear with the exciting $\mathbf{H}^{THz}$ (at relatively low excitation as we show here), the $\mathbf{E}^{THz}$-induced changes of the material refractive index are proportional to $(\mathbf{E}^{THz})^2$.

Figure 2c & 2d show the measured probe Kerr rotation under two conditions $\pm\mathbf{M} \perp \mathbf{H}^{THz}$ and $\pm\mathbf{M} \parallel \mathbf{H}^{THz}$. A giant rotation is observed with an instantaneous rise time and duration over a long delay > 20 ps. A zoom-in on the initial dynamics is shown in Fig. 2e & 2f. In the case of $\pm\mathbf{M} \perp \mathbf{H}^{THz}$, clear oscillations following the THz excitation field can be observed on the sub-ps scale superimposed on a large incoherent signal. To isolate the coherent precessions, we subtracted the trace pairs with $\pm\mathbf{B}$. The measured rotation has superimposed contributions from the excitation of both the electronic and magnetic systems. The sign of probe rotation does not depend on the sign of the THz electric

field. On the contrary, the vectorial nature of the coherent precessional dynamics suggests that the application of two stimuli with opposite signs $(\pm\mathbf{H}^{THz})$ leads to opposite torques $\mathbf{M}\times\pm\mathbf{H}_{eff}$ and thus perfectly reversed out-of-plane temporal magnetization dynamics. The probe polarization rotation direction (sign) follows the direction of this torque.

In order to distinguish between the two dynamics, we performed all the measurements in two distinct conditions with $+\mathbf{B}$ and $-\mathbf{B}$ (that is, $+\mathbf{M}$ and $-\mathbf{M}$). In this way, by subtracting the two traces the coherent precessions is obtained while the incoherent dynamics were extracted by adding the two measurements. The results are shown in Fig. 3a and 3b. In the case of $\mathbf{M}\times\pm\mathbf{H}_{eff}$, coherent magnetic precessions are excited. The Fourier analysis (Fig. 3c) unravels that they have similar spectral contents as the driving THz stimulus (Fig. 1c). A much stronger contribution to the rotation arises from the electronic excitation (nearly a factor of 9). On the contrary, in the case of $\mathbf{M}\,||\,|\mathbf{H}^{THz}|$, only the incoherent dynamics are observed (Fig. 3b). The measurement was performed by modulating the THz pump only. However, when we modulated the external bias instead, this incoherent part disappeared. This shows that the signal is not related to demagnetization and it corroborates its origin in an electronic response.

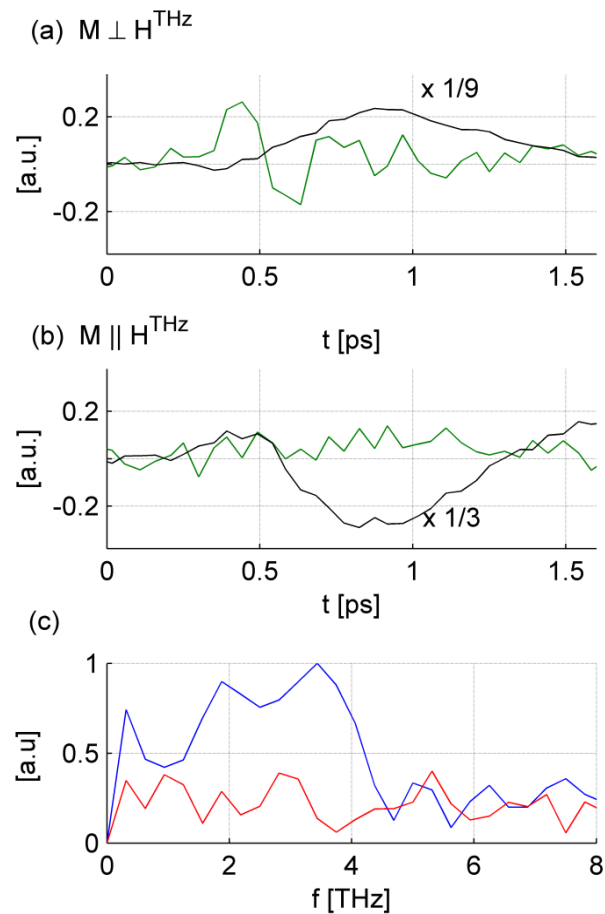

*FIG. 3 Analysis of the measurements shown in Fig.2. The coherent spin precessions (green), depending on the sign of **M**, and the electronic responses (black) are isolated from the incoherent*

*dynamics by subtracting and adding the two traces in Fig. 2e and Fig. 2f. The results are shown in (a) and (b). The green and black curves depicts the magnetic and electronic parts, respectively. In the case of $\mathbf{M} \perp \mathbf{H}^{THz}$, (a) shows clear temporal oscillations representing magnetic precessions and corresponding to the THz temporal oscillations. (c) reports the corresponding amplitude spectra to green curves in (a, blue) and (b, red).*

Finally, we would like to mention that the magnitude of the Kerr polarization rotation depends strongly on the probe initial polarization. We verified that by measuring the rotation in the case of $\mathbf{M} \,||\, |\mathbf{H}^{THz}|$ different polarization of the probe (Fig. 4). Similar to ref. 22, we obtained the maximum sensitivity for an input polarization of 45°. Nevertheless, the angle sensitivity depends on the dielectric properties of the film and the substrate.

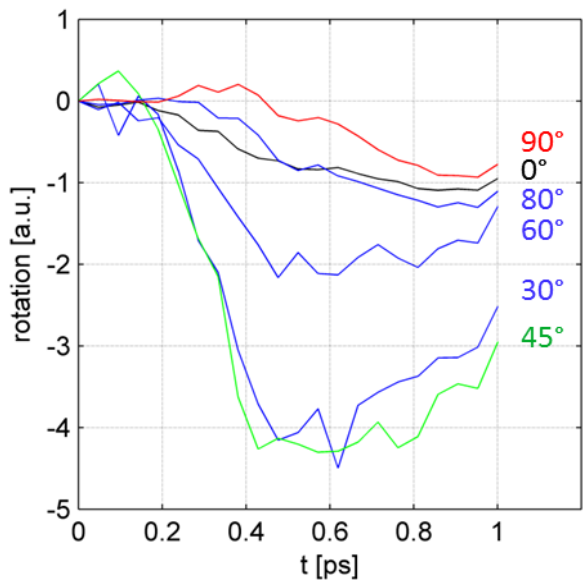

*FIG. 4. Sensitivity of the probe rotation to the polarization. 0° corresponds to P-polarization and 90° corresponds to S-polarization.*

In conclusion, we used ultra-intense THz pulses centered at 4 THz with an electric and magnetic fields of 6.7 MV/cm and 2.3 T, respectively to trigger both the electronic and magnetic response in a thin Cobalt film. In addition to the previously observed THz magnetic field-induced spin precessions, we observed electric field-driven changes in the sample dielectric tensor and thus the optical properties. The two effects are found to occur simultaneously on the ultrafast time scale of the triggering THz pulse. While the magnetic response takes place only during the THz stimulus, the electric field-induced dynamics last up to a much longer time scale (> 20 ps). We conclude that the role of the electric field can not be excluded in the studies of THz-magnetism, particularly at higher field intensities.

We gratefully thank prof. Jan Luning (Université Pierre et Marie Curie) for the sample and fruitful discussions. We are grateful to Marta Divall, Alexandre Trisorio, and Andreas Dax for supporting the operation of the Ti:sapphire laser system and the OPA. We acknowledge the support from Martin Paraliev, Rasmus Ischebeck, and Edwin Divall in DAQ. We acknowledge financial support from the Swiss National Science Foundation (SNSF) (grant no 200021_146769). MS acknowledges partial funding from the European Community's Seventh Framework Programme (FP7/2007-2013) under grant agreement n.°290605 (PSI-FELLOW/COFUND). CPH acknowledges association to NCCR-MUST and financial support from SNSF (grant no. PP00P2_150732).

[1] C. H. Back, R. Allenspach, W. Weber, S. S. P. Parkin, D. Weller, E. L. Garwin, and H. C. Siegmann, *Science* **285**, 864-867 (1999).

[2] I. Tudosa, C. Stamm, A. B. Kashuba, F. King, H. C. Siegmann, J. Stöhr, G. Ju, B. Lu,, and D. Weller, *Nature* **428**, 831-833 (2004).

[3] J. Hohlfeld, Th. Gerrits, M. Bilderbeek, Th. Rasing, H. Awano, and N. Ohta, *Phys. Rev. B* **65**, 012413 (2001).

[4] A. V. Kimel, A. Kirilyuk, P. A. Usachev, R. V. Pisarev, A. M. Balbashov, and Th. Rasing, *Nature* **435**, 655- 657 (2005).

[5] K. Vahaplar, A. M. Kalashnikova, A. V. Kimel, D. Hinzke, U. Nowak, R. Chantrell, A. Tsukamoto, A. Itoh, A. Kirilyuk, and Th. Rasing, *Phys. Rev. Lett*. **103**, 117201 (2009).

[6] E. Beaurepaire, J. C. Merle, A. Daunois, and J. Y. Bigot, *Phys. Rev. Lett.* **76**, 4250 (1996).

[7] K. Tanaka, H., Hirori, and M. Nagai, *IEEE Trans. Terahertz Sci. Technol* **1**, 301–312 (2011).

[8] H. Y. Hwang, S. Fleischer, N. C. Brandt, B. G. Perkins Jr., M. Liu, K. Fan, A. Sternbach, X. Zhang, R. D. Averitt, and K. A. Nelson, *J. Mod. Optic.* doi:doi:10.1080/09500340.2014.918200 (2014).

[9] M. Lie, H. Y. Hwang, H. Tao, A. C. Strikwerda, K. Fan, G. R. Keiser, A. J. Sternbach, K. G. West, S. Kittiwatanakul, J. Lu, *et al.*, *Nature* **487**, 345–348 (2012).

[10] R. Ulbricht, E. Hendry, J. Shan, T. F. Heinz, and M. Bonn, *Rev. Mod. Phys.* **83**, 543 (2011).

[11] T. Kampfrath, K. Tanaka, and K. A. Nelson, *Nat. Photonics* **7**, 680 (2013).

[12] O. Schubert, M. Hohenleutner, F. Langer, B. Urbanek, C. Lange, U. Huttner, D. Golde, T. Meier, M. Kira, S. W. Koch, and R. Huber, *Nat. Photonics* **8**, 119 (2014).

[13] B. Zaks, R. B. Liu, and M. S. Sherwin, *Nature (London)* **483**, 580 (2012).

[14] S. Fleischer, Y. Zhou, R. W. Field, and K. A. Nelson, *Phys. Rev. Lett.* **107**, 163603 (2011).

[15] W. R. Huang, E. A. Nanni, K. Ravi, K. –H. Hong, L. J. Wong, P. D. Keathley, A. Fallahi, L. Zapata, and F. X. Kartner, "Toward a terahertz-driven electron gun," arXiv:1409.8668

[16] T. Kampfrath, A. Sell, G. Klatt, S. Mahrlein, T. Dekorsy, M. Wolf, M. Fiebig, A. Leitensorfer, and R. Huber, *Nat. Photonics* **5**, 31–34 (2011).

[17] C. Vicario, C. Ruchert, F. Ardana-Lamas, P. M. Derlet, B. Tudu, J. Luning, and C. P. Hauri, *Nat. Photonics* **7**, 720–723 (2013).

[18] C. Vicario, B. Monoszlai, & C. P. Hauri, *Phys. Rev. Lett.* **112**, 213901 (2014).

[19] M. Shalaby and C. P. Hauri, *Nat. Commun.* **6**, 5976 (2015).

[20] M. Shalaby, M. Peccianti, D. G. Cooke, C. P. Hauri, and R. Morandotti, *Appl. Phys. Lett.* **106**, 051110 (2015).

[21] M. Shalaby, F. Vidal, M. Peccianti, R. Morandotti, F. Enderli, T. Feurer, and B. D. Patterson, *Phys. Rev. B* **88**, 140301(R) (2013).

[22] D. Allwood, P. Seem, S Basu, P. Fry, U. Gibson, and R. Cowburn, *Appl. Phys. Lett.* **92**, 2503 (2008).